\documentstyle[aps]{revtex}
\input tcilatex
\begin{document}
\title{QUANTITATIVE MODEL OF LARGE MAGNETOSTRAIN EFFECT IN FERROMAGNETIC SHAPE
MEMORY ALLOYS}
\author{A.A. Likhachev}
\address{Institute of Metal Physics, Dept. of Phase Transitions, Vernadsky St., 36,\\
252142 Kiev, Ukraine,\\
e-mail: alexl@imp.kiev.ua}
\author{K. Ullakko}
\address{Helsinki University of Technology, Dept. of Engineering Physics and\\
Mathematics, Rakentajankio 2C, 02150 Espoo, Finland,\\
e-mail: Ullakko@hut.fi}
\maketitle

\begin{abstract}
A quantitative model describing large magnetostrain effect observed in
several ferromagnetic shape memory alloys such as Ni$_2$MnGa is briefly
reported.The paper contains an exact thermodynamic consideration of the
mechanical and magnetic properties of a similar type materials. As a result,
the basic mechanical state equation including magnetic field effect is
directly derived from a general Poisson's rule. It is shown that the
magnetic field induced deformation effect is directly connected with the
strain dependence of magnetisation. A simple model of magnetisation and its
dependence on the strain is considered and applied to explain the results of
experimental study of large magnetostrain effects in Ni$_2$MnGa.
\end{abstract}

\section{Introduction}

In addition to some giant magnetostriction materials, ferromagnetic shape
memory alloys was recently suggested as a general way for the development of
a new class of the magnetic-field-controlled actuator materials \cite
{c1,c2,c3,c4,c13,c14}. It is now a goal of research projects in several
groups directed on the development of ferromagnetic alloys exhibiting also a
martensitic phase transition that would allow control of large strain effect
by application of a magnetic field at constant temperature in a martensitic
state. Numerous candidate shape memopy materials were explored including Ni$%
_2$MnGa, Co$_2$MnGa, FePt CoNi, and FeNiCoTi during past few years \cite
{c5,c6,c7}. Magnetically driven strain effect is expected to occur in these
systems. According to results reported in \cite{c8} the large strains of
0.19\% can be achieved in a magnetic field of order 8 kOe in the tetragonal
martensitic phase at 265 K of single-crystal samples of Ni$_2$MnGa. This
strain is an order of magnitude grater than the magnetostriction effect of
the parent, room temperature cubic phase.

Ni$_2$MnGa is an ordered by L21 ferromagnetic Heusler alloy having at high
temperature cubic (a = 5.822 A) crystal structure that undergoes martensitic
transformation at 276 K into a tetragonally distorted structure with
crystalline lattice parameters: a=b=5.90 A and c=5.44 A \cite{c6}. The
martensitic phase accommodates the lattice distortion connected with
transformation by formation of three twin variants twinned usually on $%
\left\langle 110\right\rangle $ planes and having the orientation of the
tetragonal symmetry axes nearly to three possible $\left[ 100\right] $
directions. The saturation value of magnetization was found to be about of
475 G. Magnetization curve of the low-temperature twinned phase usually
displays two-stage structure at 265 K \cite{c8} with a sharp crossover at
about of 1.7 kOe from easy low-field magnetization below to a hard stage
above this value up to the 8 kOe saturation field value. Such a behavior is
connected with a different response of different twin variants to the
applied field. The measurements usually show a definite magnetostrain value
along $\left[ 100\right] $ as a function of the magnetic field applied in
the same direction \cite{c8}. It is generally expected that a large
macroscopic mechanical strain induced by the magnetic field in similar type
systems is microscopically realized trough the twin boundaries motion and
redistribution of different twin variants fractions in a magnetic field. The
main thermodynamic driving forces have in this case magnetic nature and
connected with high magnetization anisotropy and differences in
magnetization free energies for different twin variants of martensite \cite
{c4,c10,c11}.

The main goal of this brief publication is to give the right thermodynamic
consideration of the mechanical and magnetic properties of a similar type
materials and represent the quantitative model describing large
magnetostrain effect observed in several ferroelastic shape memory alloys
such as Ni$_2$MnGa. It is shown that the magnetic field induced deformation
effect directly follows from the general thermodynamic rules such as Poisson
equation and connected with the strain dependence of magnetization. A simple
model of magnetization for the internally twinned martensitic state and its
dependence on the strain is considered and applied to explain the results of
experimental study of large magnetostrictive effects in Ni$_2$MnGa.

\section{General Thermodynamic Consideration}

Consider the general thermodynamic properties of the materials which can
show both the ferroelastic and the ferromagnetic properties. Most of the
shape memory alloys usually display ferroelastic behavior in martensitic
state connected with redistribution of different twin variant fractions of
martensite under the external stress applied through the motion of twin
boundaries. Ferromagnetic shape memory materials have an additional
possibility to activate the deformation process in a twinned martensitic
state by the application of magnetic field simultaneously with magnetization
of the material. According to general thermodynamic principles both the
mechanical and the magnetic properties of similar type materials can be
represented by the corresponding state equations: 
\begin{equation}
\label{e1}\sigma =\sigma \left( \varepsilon ,h\right) 
\end{equation}
\begin{equation}
\label{e1a}m=m\left( \varepsilon ,h\right) 
\end{equation}
where, the Eq.(\ref{e1}) reflects the mechanical properties through
stress-strain $\sigma -\varepsilon $ equation in presence of magnetic field $%
h$ and Eq.(\ref{e1a}) gives the magnetization value $m$ as a function of
magnetic field applied $h$ and strain $\varepsilon $. Both these equation
can be obtained from an appropriate thermodynamic potential as follows: 
\begin{equation}
\label{e2}\sigma \left( \varepsilon ,h\right) =\frac \partial {\partial
\varepsilon }\widetilde{G}\left( \varepsilon ,h\right)
\,\,\,\,\,\,\,\,\,\,\,\,\,\,\,\,\,\,m\left( \varepsilon ,h\right) =-\frac 
\partial {\partial h}\widetilde{G}\left( \varepsilon ,h\right) 
\end{equation}
where, $\widetilde{G}\left( \varepsilon ,h\right) =G\left( \varepsilon
,h\right) -hm(\varepsilon ,h),$ and $G\left( \varepsilon ,h\right) $ is the
specific Gibbs free energy at fixed temperature and pressure condition. Both
state equation are not completely independent functions and must satisfy
known Poisson's rule: 
\begin{equation}
\label{e3}\frac \partial {\partial h}\sigma \left( \varepsilon ,h\right) =-%
\frac \partial {\partial \varepsilon }m\left( \varepsilon ,h\right) 
\end{equation}
Integration of this equation over the magnetic field starting from $h=0$ at
a fixed strain gives an important representation of the mechanical state
equation including magnetic field effects: 
\begin{equation}
\label{e4}\sigma =\sigma _0\left( \varepsilon \right) -\frac \partial {%
\partial \varepsilon }\int\limits_0^hdhm\left( \varepsilon ,h\right) 
\end{equation}
According to this equation the external stress on the left is balanced in
equilibrium by both the pure mechanical stress $\sigma _0\left( \varepsilon
\right) =\sigma \left( \varepsilon ,0\right) $ resulting from the mechanical
deformation of the material at $h=0$ and the additional magnetic field
induced stress that is represented by the second term on the right in this
equation. It is also important to note that all the effect of the magnetic
field on the mechanical properties is directly determined by the strain
dependence of magnetization. In a particularly important case: $\sigma
=const=0$ one can obtain a general equation determining magnetically induced
strain (usually called as a magnetic shape memory or MSM-effect) as follows: 
\begin{equation}
\label{e4a}\sigma _0\left( \varepsilon \right) =\frac \partial {\partial
\varepsilon }\int\limits_0^hdhm\left( \varepsilon ,h\right) 
\end{equation}
and its linearized solution: 
\begin{equation}
\label{e5a}\varepsilon ^{msm}\left( h\right) =\left( \frac{d\sigma _0}{%
d\varepsilon }\right) _{\varepsilon =0}^{-1}\left( \frac \partial {\partial
\varepsilon }\int\limits_0^hdhm\left( \varepsilon ,h\right) \right)
_{\varepsilon =0} 
\end{equation}
that can be used when $\varepsilon $ is much less than a martensite lattice
tetragonal distortion value $\varepsilon _0$. According to Eqns.(\ref{e4a})
and (\ref{e5a}) the magnetization and its dependence on the strain is
responsible for the MSM-effect and is the main subject for detailed
discussion and modeling.

\section{The model and its application to Ni$_2$MnGa tetragonal martensite}

Consider a typical situation corresponding to measurements of large strain
induced by the magnetic field in the tetragonal internally twinned
martensite of Ni$_2$MnGa obtained from the austenitic single crystal studied
in \cite{c8} {}when the magnetic field is applied along $\left[ 100\right] $
direction of parent austenitic phase and strain measurements were performed
in the same axial direction. In this case the crystallographic $\left[
100\right] $ $\left[ 010\right] $ $\left[ 001\right] $ axes for all three
possible twin variants of the tetragonal martensitic phase will be nearly
parallel to the external field applied. More exactly, the crystallographic
orientation relationships between the austenitic and martensitic phases can
be obtained by using the usual methods of the crystallographic theory.
Additional small rotations of the tetragonal phase axes are expected but the
corresponding rotation angles can not exceed few degrees in the case of Ni$%
_2 $MnGa and may be neglected for simplicity. Fig.1. schematically shows the
expected alignment of the magnetic field applied, crystallographic
orientations and magnetization curves for three possible tetragonal phase
variants.

\FRAME{ftbhxF}{4.688in}{3.428in}{0.778in}{Schematic representation of the %
magnetic field alignment and magnetisation behavior for three different %
crystallographic variants of the tetragonal martensitic phase.}{}{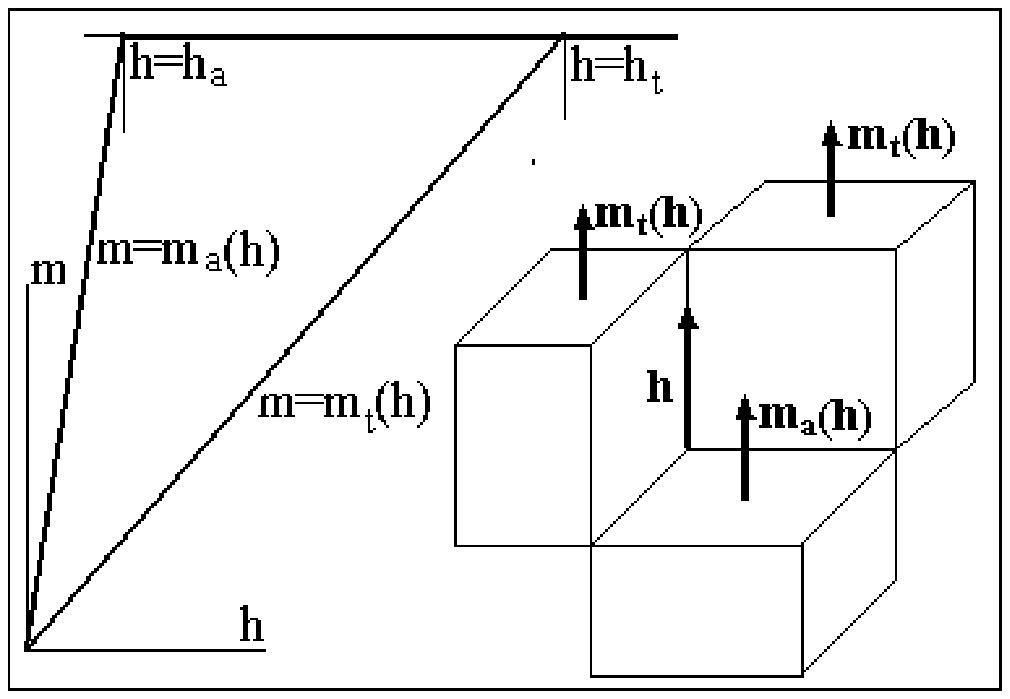}%
{\GRAPHICSPS{\FILENAME{likhfg1.EPS}}}
Therefore, the magnetic field is applied along the tetragonal symmetry axis
only for one type twin variant (which is called here the axial, or $a$-type)
and simultaneously in the transversal direction in respect to the tetragonal
symmetry axes of another two (transversal, or $t$-type) variants. The
investigation of magnetization properties of Ni$_2$MnGa performed for a
single tetragonal variant of martensite obtained by the mechanical
compression method \cite{c12} has shown a considerable difference between
the magnetization curves along the tetragonal $\left[ 100\right] $ direction
in comparison to another transversal $\left[ 010\right] $ and $\left[
001\right] $ directions. It was found that a tetragonal axis is the easiest
magnetization direction and requires considerably less value of saturation
field $h_a$ than a saturation field $h_t$ characterizing magnetization in
two hard transversal directions as it is schematically shown in Fig.1. In a
general case the calculation of magnetization for the material with a
complicated twin microstructure geometry requires some special approach. In
this paper we will ignore, for simplicity, a similar type problem and will
consider these effects in other publications. Taking into account the
presence of magnetic anisotropy and difference in magnetization behavior
between the axial $m_a(h)$ and transversal $m_t(h)$ twin variants we
consider a simple model of magnetization for the multi-variant martensitic
state that gives the main contribution into magnetisation insensitive to the
fine details of twin microstructure. This model treats the multi-twin
martensitic state as a composite material consisting of an easy
magnetization area occupied by axial type twins and hard magnetization
region of two transversal twin variants. Denoting as $x$ the total volume
fraction of the axial twin domain and $(1-x)$ - transversal type twin domain
fractions, respectively, one can write the magnetization of the material as
follows: 
\begin{equation}
\label{e6}m\left( x,h\right) =xm_a(h)+(1-x)m_t(h) 
\end{equation}
where, $m_a(h)$ and $m_t(h)$ are specific magnetization functions for the
axial and transversal variants, respectively. On the other hand, the
macroscopic strain along the axial direction can be found from a similar
type equation: 
\begin{equation}
\label{e7}\varepsilon =x\varepsilon _a^0+(1-x)\varepsilon _t^0=\frac 32%
\varepsilon _0(x-\frac 13) 
\end{equation}
where, the diagonal matrix elements $\varepsilon _a^0=\varepsilon _0$ and $%
\varepsilon _t^0=-\frac 12\varepsilon _0$ represent the relative tetragonal
distortion of the martensite crystal lattice along its tetragonal axis and
two transversal directions, respectively. The compression distortion $%
\varepsilon _0=5.4\%$ along the tetragonal symmetry axis was found in case
of Ni$_2$MnGa. One can easily exclude the fraction dependence from these two
equations and obtain the magnetization as function of the macroscopic strain
for the internally twinned martensitic state: 
\begin{equation}
\label{e8}m\left( \varepsilon ,h\right) =\left\{ \frac 13m_a(h)+\frac 23%
m_t(h)\right\} +\frac 23\left( \varepsilon /\varepsilon _0\right) \left\{
m_a(h)-m_t(h)\right\} 
\end{equation}
This equation immediately reproduces all the main peculiarities of the
experimental magnetization curve including the sharp change of its slope at $%
h=1.75\,\,\,kOe$ as indicated in Fig.2. This singularity exactly appears at $%
h=h_a$ where the easy stage of magnetization process inside of the axial
twin variants domain is finished. According to Eq.(\ref{e8}) $m\left(
\varepsilon _0,h\right) =m_a\left( h\right) $ and $m\left( -\varepsilon
_0/2,h\right) =m_t\left( h\right) $ so, one can use this fact to obtain both
the $h_a=1.75\,\,\,kOe$ and $h=8\,\,\,kOe$ from the experimental
magnetization curves measured in the multi-variant state. The model
magnetization curve $m\left( 0,h\right) $ corresponding to zero strain value
shows the same type behavior and singularity in slope as the experimental
one. The difference between them is caused by the second term in Eq.(\ref{e8}%
) that gives an additional strain dependent contribution into the
magnetisation. This contribution is directly connected with the MSM-effect
and can be easily taken into account just after its calculation.

\FRAME{fbhxF}{298pt}{300pt}{0pt}{Experimental and model magnetisation %
curves that show the compression and tensile strain effect on the macroscopic %
magnetisation of martensitic phase caused by the relative change of the twin %
variant fractions.}{}{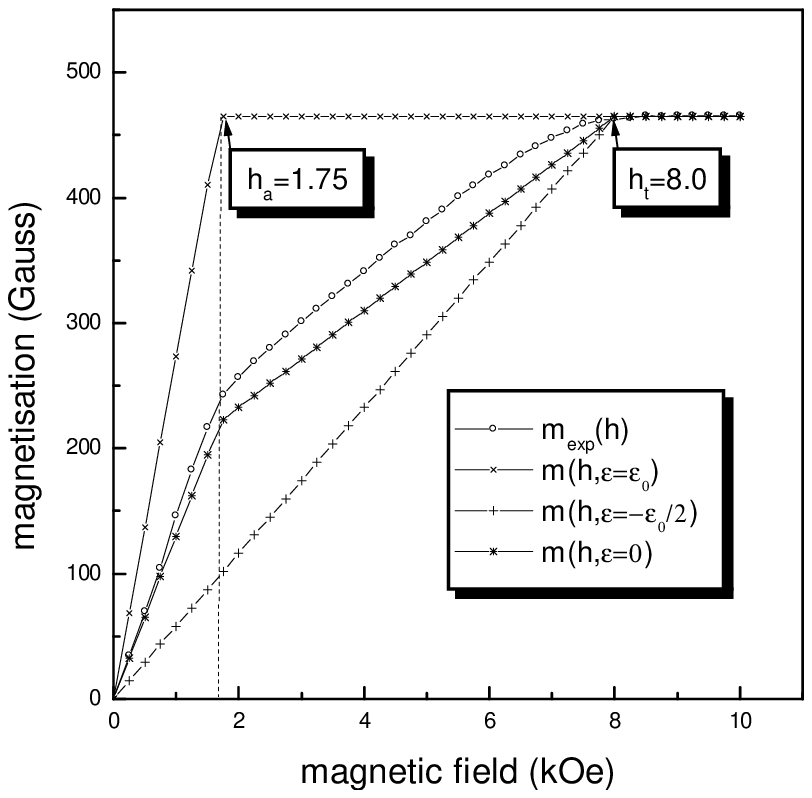}{\GRAPHICSPS{\FILENAME{LIKHFG2.EPS}}}
One can also obtain the final equations representing the effect of magnetic
field on the strain by using basic expressions (\ref{e5a}) derived before
from the general thermodynamic consideration: 
\begin{equation}
\label{e10}\varepsilon ^{msm}\left( h\right) =\frac 23\left( \varepsilon _0%
\frac{d\sigma _0}{d\varepsilon }\right) _{\varepsilon
=0}^{-1}\int\limits_0^hdh\left\{ m_a(h)-m_t(h)\right\} 
\end{equation}

\section{Discussion and conclusions}

As follows from this equation, two factors determine the strain value and
its field dependence. The first one is proportional to the slope of
stress-strain curve and can be found from the usual mechanical compression
test without magnetic field applied. The integral term reflects the effects
of magnetic anisotropy and determines the functional magnetic field
dependence of the strain. In particular, in absence of the magnetization
anisotropy when $m_a(h)=m_t(h)$ deformation effect is also vanishes. The
saturation level of the strain is achieved at $h=h_t$ and above where $%
m_a(h)=m_t(h)=m^{sat}$ and where the material has its maximal magnetization
value $m^{sat}$. One can easily obtain the corresponding saturation value of
the strain performing the necessary integrations in Eq.(\ref{e10}) as
follows: 
\begin{equation}
\label{e11}\varepsilon _{sat}^{msm}=\frac 13\left( \varepsilon _0\frac{%
d\sigma _0}{d\varepsilon }\right) _{\varepsilon =0}^{-1}\left(
h_t-h_a\right) m^{sat} 
\end{equation}
Precise quantitative estimation of the saturation strain requires, in
general, the corresponding mechanical testing. Here, we will use a simple
estimation of $d\sigma _0/d\varepsilon \sim \sigma _0/\varepsilon _0$. So, $%
\varepsilon _{sat}^{msm}\sim \frac 13\left( \sigma _0\right) ^{-1}\left(
h_t-h_a\right) m^{sat}$ where the characteristic stress $\sigma _0$
representing ferroelastic mechanical behavior of the material is expected to
be about of 20MPa in Ni$_2$MnGa martensite. Using also the values of $%
h_t\sim 8kOe$ , $h_a\sim 1.75kOe$ and $m^{sat}\sim 475G$ found from the
magnetization curve analysis one can obtain a simple estimation: $%
\varepsilon _{sat}^{msm}\sim 0.49\%.$ More precise estimation that follows
from the mechanical testing results gives $d\sigma _0/d\varepsilon \sim
\left( 2\div 3\right) \sigma _0/\varepsilon _0.$ Consequently, $\varepsilon
_{sat}^{msm}\sim (0.24\div 0.16)\%$ which is in a better quantitative
agreement with $\varepsilon _{sat}^{msm}\sim 0.14\%$ experimental value. In
order to achieve the larger magnetostrain effect comparable with the lattice
tetragonal distortion value $\varepsilon _0\sim 5\%$ one will need materials
with very low $\sigma _0\sim 2MPa$ detwinning stress value. This task seems
can be considered as the realistic one because the observation of $\sigma
_0\sim 2MPa$ \cite{c6} and $\sigma _0\sim 8MPa$ \cite{c10} were reported in
some publications.

Fig.3. shows the field behavior of the strain that follows from the model
and its change for the different values of the magnetic anisotropy factor $%
k=h_a/h_t$ defined as a ratio between the axial and transversal saturation
fields.

\FRAME{ftbhxF}{300pt}{256pt}{0pt}{Magnetic anisotropy effect on the strain %
vs. magnetic field behavior according to the model calculations. }{}{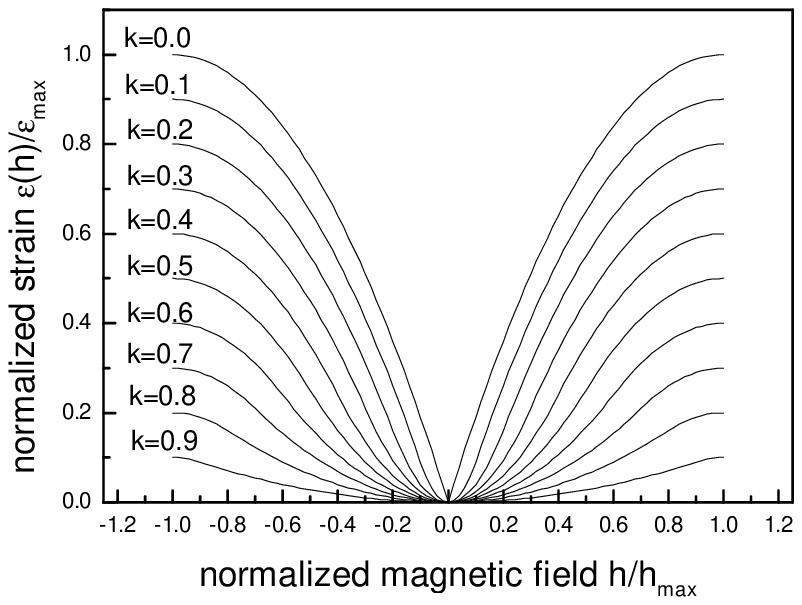}%
{\GRAPHICSPS{\FILENAME{LIKHFG3.EPS}}}

The dimensionless strain response $\varepsilon ^{msm}\left( h\right)
/\varepsilon _{\max }$ normalized by, 
\begin{equation}
\label{e12}\varepsilon _{\max }=\frac 13\left( \varepsilon _0\frac{d\sigma _0%
}{d\varepsilon }\right) _{\varepsilon =0}^{-1}h_tm^{sat} 
\end{equation}
increases from zero value at $k=1$ simultaneously with a corresponding shape
change and shows the maximal possible deformation effect and linear type
singularity for the low field strain behavior at $k=0$. This case
corresponds to the maximally strong anisotropy when the axial saturation
field becomes infinitely small $h_a\longrightarrow 0$ and $m_a(h)$
immediately achieves its saturation level $m^{sat}$ starting from an
arbitrary low magnetic field and then still remains equal to a constant $%
m^{sat}$ value during the magnetization process. Therefore, one can conclude
that a linear low field behavior usually predicted in some previously
developed models \cite{c10,c11} is directly connected with their assumption
on the complete saturation of magnetization for the axial type twin
variants. According to the present model such a type of assumption can be
physically reasonable in the limit $h_a\longrightarrow 0$ only. In other
case the strain shows the normal parabolic type behavior in the low field $%
h<h_a$ region in agreement with the experimental observations. A good
correspondence between the model and experimental results is indicated in
Fig.4. We neglected here the small hysteresis effects which are usually
observed assuming to give the more detailed discussion of this problem by
using some new developments and quantitative descriptions of hysteresis in
shape memory materials.

\FRAME{fthxF}{300pt}{258pt}{0pt}{Magnetostrain effect: comparison results %
between the model and the experiment}{}{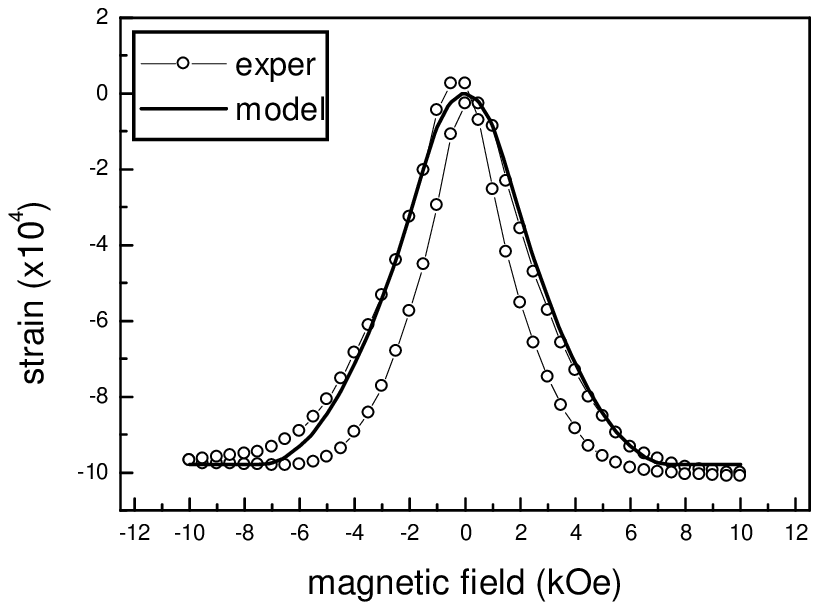}{\GRAPHICSPS{\FILENAME{%
LIKHFG4.EPS}}}
{\bf Acknowledgments}

Authors acknowledge the Physical and Mat. Science Departments of the
Helsinki Technological University supported this work.

\end{document}